\documentstyle[12pt]{article}

\thicklines                         
\def\d{\delta}

\def\k{\kappa}

\def\m{\mu}
\def\n{\nu}

\def\x{\xi}

\def\D{\Delta}

\def\cg{{\cal G}}

\def\cs{{\cal S}}

\def\cv{{\cal V}}

\def\cbo{{\,\raise-.15ex\Sc [\,}}                       
 
\def\ddt#1{{\buildrel {\hbox{\LARGE .\kern-2pt.}} \over {#1}}}

\def\beq{\begin{equation}}
\def\eeq{\end{equation}}
\def\bqry{\begin{eqnarray}}
\def\eqry{\end{eqnarray}}

\def\beqn#1{ \renewcommand{\theequation}{#1} 
             \begin{eqnarray} }
\def\eeqn{ \renewcommand{\theequation}{\arabic{equation}}
           \end{eqnarray} }

\def\beqr#1{ \setcounter{equation}{#1} 
             \begin{eqnarray} }

\def\eeqr{\end{eqnarray}}

\def\beqrabc#1{ \setcounter{equation}{0}
                \renewcommand{\theequation}{#1\alph{equation}} 
                \begin{eqnarray} }
\def\beqrn#1#2{ \setcounter{equation}{#2}
                \renewcommand{\theequation}{#1.\arabic{equation}} 
                \begin{eqnarray} }
\def\seeq#1{eq.~(\ref{#1})}

\def\seeqs#1{eqs.~(\ref{#1})}

\def\seneq#1{~(\ref{#1})}

\def\JMP#1{Jour. Math. Phys. {\bf #1}}
\def\NPB#1{Nucl. Phys. {\bf B#1}}
\def\NPBP#1{Nucl. Phys. (Proc. Suppl.) {\bf B#1}}
\def\PLB#1{Phys. Lett. {\bf B#1}}
\def\PRD#1{Phys. Rev. {\bf D#1}}

\def\PRL#1{Phys. Rev. Lett. {\bf #1}}

\def\sstyle{\scriptstyle}

\def\frac#1#2{ {\sstyle {#1\over #2} } }

\newcommand{\eq}{\ref}

\def\tk{\tilde\k}

\begin{document}
\hyphenation{fer-mio-nic per-tur-ba-tive}

\noindent March 1998 \hfill TAUP--2481--98  \\
\noindent \phantom{February 1998} \hfill Wash. U. HEP/98-60 \\
\noindent \phantom{February 1998} \hfill HU-EP-98/20 \\

\begin{center}
\vspace{15mm}
{\large\bf More on Lattice BRST Invariance}
\\[15mm]
Wolfgang Bock$^a$, Maarten F.L. Golterman$^b$
and Yigal Shamir$^c$
\\[5mm]
$^a$Institute of Physics, Humboldt University \\
Invalidenstr.~110,~10115~Berlin,~Germany
\\[2mm]
$^b$Department of Physics, Washington University\\
St.~Louis,~MO~63130,~USA
\\[2mm]
$^c$School of Physics and Astronomy, Tel-Aviv University \\
Ramat~Aviv,~69978~Israel
\\[5mm]
{ABSTRACT}
\\[2mm]
\end{center}

\begin{quotation}
In the gauge-fixing approach to (chiral) lattice gauge theories, 
the action in the U(1) case implicitly contains a {\it free} ghost term, 
in accordance with the continuum abelian theory.
On the lattice there is no BRST symmetry and,
without fermions, the partition function is strictly positive.
Recently, Neuberger pointed out in hep-lat/9801029 that a different choice of
the ghost term would lead to a BRST-invariant lattice model,
which is ill-defined nonperturbatively. 
We show that  
such a lattice model is inconsistent already in perturbation theory,
and  clearly different from the gauge-fixing approach.
\end{quotation}

\newpage

\noindent {\it 1.} A  central difficulty
in constructing chiral lattice gauge theories stems from the coupling
between the fermions and the longitudinal degrees of freedom
of the lattice gauge field. The physical reason for this coupling
is the need to correctly reproduce the contribution to the anomaly
for each fermion species, in the background of smooth gauge fields~\cite{lkjs}.
However, for generic lattice gauge fields,
this (momentum-dependent) coupling is not small,  and renders the
fermion spectrum vector-like instead of chiral in many 
proposals for lattice chiral gauge theories~\cite{rev}.

Gauge fixing is a natural way to try 
control the dynamics of the longitudinal
degrees of freedom~\cite{roma,sml,mgys}. 
In the gauge-fixing approach one transcribes to
the lattice a chiral gauge theory, whose action in the continuum 
already contains
Lorentz gauge-fixing and ghost terms.
One then looks for a continuum limit
governed by renormalized perturbation theory, requiring that
 this continuum limit indeed corresponds to the 
target gauge-fixed chiral gauge theory.

In the U(1) case, strong evidence 
for the existence of this continuum limit 
was found by us in a ``reduced'' model, 
where one restricts the U(1) gauge field to the trivial orbit. 
(In the future we plan to investigate the model
with fully dynamical U(1) gauge fields.) 
The continuum limit corresponds to a continuous phase transition
between a normal broken phase and an exotic broken phase where, 
in addition, rotation symmetry is broken by a vector condensate~\cite{sml}.
Analytical and numerical evidence 
for the existence and continuity of the phase transition
is given in refs.~\cite{bgs1,bgs3}. 
Evidence that the correct chiral 
spectrum is obtained in the continuum limit
is given in refs.~\cite{bgs3,bgs2}. As explained in ref.~\cite{LAT97}
(which  contains a less technical  account  of our work) 
the gauge-fixing approach does not contradict the Nielsen-Ninomiya
theorem~\cite{NN,lkjs} as reformulated for interacting lattice theories in
ref.~\cite{ysnogo}.

The lattice action of the gauge-fixing approach
(\seeq{S} below) includes a {\it free} ghost
term in the U(1) case, in accordance with the target continuum abelian
theory. Evidently, this exactly decoupled ghost sector does not 
affect any observable constructed out of the vector and fermion fields,
hence it was dropped from the definition of the U(1) lattice action
given in refs.~[5-8] (for a concise formulation see ref.~\cite{bgs2}). 
With the ghost action included, 
one can formulate BRST transformations, but the lattice action is not
BRST invariant. Following ref.~\cite{roma},
one adds counterterms to the action in order to restore
BRST invariance in the continuum limit. In particular, 
the continuous phase transition
mentioned above corresponds to a vanishing photon mass.

\newpage

\noindent {\it 2.} 
Recently~\cite{hn}, Neuberger pointed out that a different choice 
of the ghost action exists, such that the sum of the gauge-fixing
term of refs.~[5-8] and the new ghost term is BRST invariant.
The U(1) lattice model defined using that BRST invariant 
action is in fact ill-defined.
This is a consequence of a general ``no-go'' theorem~\cite{hnogo},
also due to Neuberger, which asserts 
that the partition function itself, as well as 
(unnormalized) expectation values
of gauge-invariant operators, vanish identically in
a lattice model with exact BRST invariance. 
As a result, (normalized) expectation 
values of gauge-invariant operators
always lead to ill-defined expressions of the form  ``$0/0$''.

We will first describe Neuberger's observation in detail,
and then explain why it is irrelevant for the gauge-fixing approach.
The BRST-invariant U(1) model which was considered in ref.~\cite{hn}
is defined by the  path integral
\bqry
  Z \!\!\! &=& \!\!\! \int {\cal D} U \; {\cal D}  \bar{c} 
  \;  {\cal D} c 
  \; \exp(-S_{\rm BRST}(U;\bar{c},c))\,, \label{Z} \\
  S_{\rm BRST}(U;\bar{c},c) \!\!\! &=& \!\!\! 
  S_{\rm gaugeinv}(U) +  S_{\rm gaugefix}(U) 
  + S_{\rm ghost}(U;\bar{c},c) \,.
\label{S'}
\eqry 
This model contains vector and ghost fields, but no matter fields.
The gauge-invariant term in the action 
represents the standard plaquette action, 
which in the classical continuum limit reduces to   
${1\over 4}\int d^4x\, \sum_{\m \n}F_{\m\n}^2$. The gauge-fixing
term has the general form 
\beq
  S_{\rm gaugefix}(U) = {1\over 2 \x} \sum_x \cg_x(U)^2 \;,
\label{SxGx}
\eeq
where $\xi>0$ is the gauge-fixing parameter. $\cg_x(U)$, which we will call
the gauge condition, is a  real local functional of 
the lattice link variables $U_{x\m}=\exp(igA_{x\m})$, 
which is continuously differentiable
over the (compact) space of U(1) lattice gauge-field configurations.
The general form of the ghost term is
\beq
S_{\rm ghost}(U;\bar{c},c) = \sum_{xy} \bar{c}_x \Omega_{xy}(U) c_y \;,
\label{Sgh}
\eeq
where $c$ and $\bar{c}$ are ghost and anti-ghost fields. The ghost
operator is 
\beq
  \Omega_{xy}(U) = \sum_\m \; 
  {\d \cg_x(U)\over \d A_{y\m}} \; \D^+_{y\m} \;,
\label{opr}
\eeq
where $\D^+_{y\m}$ is the forward lattice derivative,
defined as $\D^+_{x\m}f=f_{x+\hat\m}-f_x$ for any function $f_x$.
The model in eq.~(\eq{Z}) has an exact 
BRST invariance if the same $\cg_x(U)$ enters
both the gauge-fixing and ghost terms. 
As mentioned above, in this case it was proved 
by Neuberger that the partition function (\eq{Z}) itself, as well 
as (unnormalized) expectation values of gauge invariant
operators, vanish \cite{hnogo}. 

In ref.~\cite{hn} Neuberger showed that 
a BRST invariant action exists whose gauge-fixing term\seneq{SxGx}
coincides with the one defined in refs.~[5-8] up to a  trivial constant
$\cv M$ where $\cv$ is the lattice volume.
The gauge-fixing term advocated in refs.~[5-8] has the form   
\beq
   S_{\rm gaugefix}^{\rm L}(U) = \tk \left\{               
   \sum_{xyz} \; \Box_{xy}(U) \Box_{yz}(U)  - \sum_x B_x^2(U) \right\}
\;,\quad\quad   \tk={1\over 2 \x g^2 } \;,
\label{Sgf}
\eeq
where
\beq
  B_x(U)={1\over 4} \; \sum_\m ( V_{x-\hat\m,\m} + V_{x\m} )^2\;, 
\label{B}
\eeq
\beq
  V_{x\m} =  \mbox{Im}\; U_{x\m} = \sin(gA_{x\m}) \;,
\label{V}
\eeq  
and $\Box_{xy}(U)=
\sum_\m(\d_{x+\hat\m,y}U_{x\m}+\d_{x-\hat\m,y}U_{y\m}^\dagger)
-8\d_{x,y}$ is the covariant nearest-neighbor lattice laplacian. 
In the classical continuum limit $S_{\rm gaugefix}^{\rm L}(U)$ reduces 
to the Lorentz gauge-fixing action, 
$\frac{1}{2\x} \int d^4 x (\sum_\m \partial_\m A_\m )^2$.
The other properties of $S_{\rm gaugefix}^{\rm L}(U)$ are  
summarized in Sect.~3. 
Now, one can write
\beq
S_{\rm gaugefix}^{\rm L}(U) = {1\over 2\x} \sum_x \cs_x(U) \;.
\label{Sx}
\eeq
The  BRST invariant action is defined by picking~\cite{hn}
\beq
\cg_x(U) = \sqrt{ \cs_x(U) +M }   \;, 
\label{GFHN}
\eeq
where $M$ is a constant chosen such that  $M > -{\rm min} \{\cs_x(U)\}$.
Note that the range of the functional $\cs_x(U)$ 
over the entire lattice configuration
space is a bounded closed interval, hence ${\rm min} \{\cs_x(U)\}$
is necessarily finite. (In fact, ${\rm min} \{\cs_x(U)\}=0$ \cite{mgys}.)
As a special case of Neuberger's theorem,
the partition function\seneq{Z} vanishes
if the functional\seneq{GFHN} is used in its definition.

The gauge-fixing approach evades this inconsistency by {\it not}
having BRST symmetry on the lattice.
In the absence of fermions, the Boltzmann weight
of the gauge-fixing approach in the U(1) case is strictly positive 
(see Sect.~3), which implies that the ``$0/0$'' problem does not occur.     
 Moreover, we wish to demonstrate that {\it perturbative} consistency
already excludes the ghost action constructed in ref.~\cite{hn}.

Perturbation theory is an expansion around a
classical vacuum, {\it i.e.} a translationally invariant
global minimum of $S_{\rm gaugefix}(U)$ on the trivial orbit.
We consider in the following 
a gauge condition $\cg_x(U)$  with a strictly 
positive range, {\it i.e.} $\cg_x(U) >0$ and 
which is translationally covariant, 
{\it i.e.} $\cg_x(U_{y \m})=\cg_{x-z}(U_{y-z,\m})$. 
An example is the gauge condition\seneq{GFHN}.
We will prove now that for such a gauge condition 
the Faddeev-Popov operator is identically zero, {\it i.e.} $\Omega_{xy}=0$,
on a classical vacuum. 

The proof is very simple. Let $U^0_{z\m}=\exp(igA^0_{z\m})=U^0_{\m}$ be 
a translationally invariant saddle point of $S_{\rm gaugefix}(U)$. Then
\beq   
  \cg_x(U^0_{\m})  \,\Omega_{xy}(U^0_{\m})
   = {1\over 2} \; \sum_\n \left.{\d \cg_x^2 \over \d A_{y\n}} 
   \right|_{U=U^0_{\m}}
   \; \D^+_{y\n} 
   = 0 \;.
 \label{proof}
 \eeq   
The first equality follows from \seeq{opr}. 
The second equality follows because,
by \seeq{SxGx},
a translationally invariant $U^0_{\m}$ 
is a saddle point of 
$S_{\rm gaugefix}(U)$ if and only if it is a saddle point of 
$\cg_x^2(U)$ for any $x$.
Notice now that $\cg_x(U^0_\m)\ne 0$ by assumption.
Dividing both sides of \seeq{proof} by $\cg_x(U^0_\m)$,
we obtain $\Omega_{xy}(U^0_\m)=0$.

The conclusion is that perturbation theory
is undefined if $\cg_x(U)$ is a  strictly positive functional, 
since the tree-level ghost operator $\Omega(U^0_\m)$
vanishes identically.  We note that the gauge condition,
eq.~(\ref{GFHN}), is completely determined by the requirement that the
gauge-fixing term, eq.~(\ref{SxGx}), of the BRST-invariant action should
coincide (up to the constant ${\cal V}M$) with $S^{\rm L}_{\rm
gaugefix}(U)$. 
Hence, this also proves that $S_{\rm gaugefix}^{\rm L}(U)$ 
cannot be the gauge-fixing term of any BRST invariant action that has the 
correct classical continuum limit. (Recall that, for the Lorentz gauge, 
the  quadratic part of the 
continuum ghost action is $\bar{c}\,\Box\, c$, and not zero,
in abelian as well as in nonabelian theories.)

As was shown in ref.~\cite{luescher}, if one is interested {\it only} in 
perturbation theory, one can employ the BRST construction just as in
the continuum. Of course, one has to make sure that the gauge-fixing
and ghost terms both have the correct classical continuum limit.
In view of the above result, this implies that one must use an 
{\it indefinite-sign} functional  for $\cg_x(U)$. 
We conclude this section with an example  of this.
Consider the lattice discretization $\cg_x^{\rm L,naive}(U)$ of
 the Lorentz gauge condition $\sum_\m \partial_\m  A_\m $, with

\beq
  \cg_x^{\rm L,naive}(U) = {1\over g}\sum_\m \D^-_{x\m} V_\m \;,
\label{Gcov}
\eeq
where $\D^-_{x\m}$ designates the 
backward lattice derivative, and
$V_\m$ is defined in \seeq{V}. 
One expects that the
equation $\sum_\m \D^-_{x\m}[\sin(gA_{\m}-\D_{\m}^+\theta)] = g v_x$
can be solved for sufficiently small $A_\m$ and $v$.
Therefore the range of $\cg_x^{\rm L,naive}(U)$ 
contains an open neighborhood of zero and   
$\cg_x^{\rm L,naive}(U)$ is an indefinite-sign functional.  
Eq.~(\eq{proof}) is now fulfilled on a classical vacuum 
because  $\cg_x^{\rm L,naive}(U^0_\m)=0$ for all $x$.
Since furthermore the gauge-fixing action  
\beq
  S_{\rm gaugefix}^{\rm L,naive}(U)=
  {1\over 2\x}\sum_x(\cg_x^{\rm L,naive}(U))^2 \;,
\label{Snaive}
\eeq
and the Faddeev-Popov operator have 
the correct classical continuum limit,  $\cg_x^{\rm L,naive}(U)$ is a 
consistent gauge condition at the level of perturbation theory.

\vspace{4ex}

\noindent {\it 3.} 
In this section we discuss the gauge-fixing approach in some more detail. 
Specifically, we will consider the lattice transcription 
of a Lorentz gauge-fixed U(1) theory, where the continuum theory consists of
free photons only. Due to the presence of a quadratic
covariant gauge-fixing term we expect to get all four polarizations
as free, uncoupled states in the continuum limit of the lattice model.
(We emphasize that the question here 
is not the practicality of working with
a gauge-fixed U(1) lattice theory, but, rather, its existence.)
The lattice model is now defined by the action
\beq
  S(U;\bar{c},c) = S_{\rm gaugeinv}(U) +  S_{\rm gaugefix}^{\rm L}(U) 
  + S_{\rm ghost}^{\rm L}(\bar{c},c) + S_{\rm counterterm}(U) \,.
\label{S}
\eeq
The gauge-invariant term is again the plaquette action.
$S_{\rm gaugefix}^{\rm L}(U)$ is the lattice discretization 
of the Lorentz gauge-fixing action  introduced in eq.~(\eq{Sgf}).
The {\it free} ghost action is
\beq
  S_{\rm ghost}^{\rm L}(\bar{c},c) = 
  \sum_{xy}  \bar{c}_x \{ -\Box_{xy} + \m^2 \,\d_{xy} \} c_y  \,,
\label{SghostL}
\eeq
where for definiteness we have chosen $\Box_{xy}$ as the nearest-neighbor
free lattice laplacian.
We have added an infinitesimal mass term ($0 < \m^2\ll 1$) to avoid
the trivial finite-volume zero mode. One can safely set $\m=0$
after the infinite volume limit is taken. 
(Alternatively, one could {\it e.g.} choose antiperiodic boundary conditions.)
It is evident from \seeqs{S} and\seneq{SghostL} that the Boltzmann
weight of the gauge-fixing approach is strictly positive in the U(1) case.

Given the U(1) action\seneq{S}, one can formulate lattice BRST transformations,
but obviously, $S(U;\bar{c},c)$ is not BRST invariant.  
Following the procedure proposed and outlined in ref.~\cite{roma}
(see in particular section 6 of that paper),
one adds counterterms to the action, in order to restore
BRST invariance in the continuum limit.  In perturbation theory,
this means that the  continuum limit
of any  correlation function should obey the relevant {\it continuum}
BRST identity. Because the ghosts are
free, it is possible to impose BRST invariance without ghost
counterterms, since all connected ghost correlation functions agree with
the continuum ones in the continuum limit already. 
As  we already mentioned in the introduction, the decoupled ghost sector
cancels out from the expectation value of any operator constructed from
the gauge (and/or matter) fields, hence it was dropped in refs.~[5-8].
(The U(1) continuum action is BRST invariant also with massive
photon and ghost fields, provided their masses are equal 
(see for instance, ref.~\cite{collins}).
On the lattice, one can impose the BRST identities
of the massive theory in the continuum limit, 
sending $\m\to 0$ in the end. 
Yet another possibility is to use the action {\it without} the free
ghost term, in which case it is strictly speaking more appropriate 
to talk about recovering Ward identities rather than BRST invariance
in the continuum limit.) 

The gauge-fixing action density ({\it cf.} \seeq{Sx})
can be written as 
$\cs_x(U) = \cs_x^{(1)}(U) + \cs_x^{(2)}(U)$, where 
$\cs_x^{(1)}(U) = (\cg_x^{\rm L,naive}(U))^2$,  {\it cf.} \seeq{Gcov}. 
Thus, $\cs_x^{(1)}(U)$ corresponds 
to the naive lattice transcription of the 
continuum $(\sum_\m \partial_\m A_\m )^2$ 
discussed in the previous section.
While perturbation theory is self-consistent in this case, 
it may be unreliable in view of the proliferation of lattice Gribov
copies of the $U_{x\m}=1$ classical vacuum for the gauge condition 
$\cg_x^{\rm L,naive}(U)$, each of which is a global minimum of 
$S_{\rm gaugefix}^{\rm L,naive}(U)$ \cite{sml} (see also ref.~\cite{copies}). 
In particular, the existence of the
continuous phase transition where we want to take the continuum limit
is {\it a priori} not guaranteed. This is remedied by the addition 
of $\cs_x^{(2)}(U)$. The latter contains
only irrelevant operators, and has a unique absolute minimum
at $U_{x\m}=1$. (That irrelevant terms can have a profound effect
on the continuum limit should not come as a surprise, as the 
example of the Wilson term for lattice Wilson fermions shows.) 
We now summarize the key properties of the action $S(U;\bar c,c)$,
eq.~(\ref{S}),
starting with the results of Sect.~2:
\begin{itemize}

\item  $S(U;\bar c,c)$ is not
      invariant under BRST transformations. Moreover,
      there does not exist a BRST invariant lattice action
      with the correct classical continuum limit,
      whose gauge-fixing term coincides with $S_{\rm gaugefix}^{\rm L}(U)$. 

\item $S_{\rm gaugefix}^{\rm L}(U)$ has a unique absolute minimum
      at $U_{x\m}=1$~\cite{mgys}.

\item  $S(U;\bar c,c)$ has the correct classical continuum limit. 

\end{itemize}
The second property ensures that the euclidean functional integration 
is dominated by the unique global maximum of the Boltzmann weight.
 The third property implies that kinetic terms exist for all
polarizations of the gauge field as well as the ghost fields.
Therefore, perturbation theory
is well-defined and renormalizable. This  is at the heart of 
the good agreement between one-loop perturbation theory
and nonperturbative numerical  results found in the 
reduced model~\cite{bgs1,bgs2}.

As explained above, in order to recover
BRST invariance, we have introduced 
in \seeq{S} a finite number of counterterms
that correspond to all relevant and marginal operators
which are allowed by the exact lattice symmetries~\cite{roma}.
The only dimension-two counterterm
is the photon mass term 
\beq
  S_{\rm mass}(U) = -2\k \sum_{x\m} \mbox{Re}\; U_{x\m} \;.
\label{ct}
\eeq
So far, this is the only counterterm that we have studied
in detail~[5-8]. 
The mass counterterm is crucial because the continuum limit
mentioned in the introduction corresponds to a vanishing photon mass.
This is achieved by tuning $\k$ in \seeq{ct} to its critical value.
A brief discussion of nonderivative dimension-four
counterterms is given in ref.~\cite{mgys}. (In the future we plan 
to investigate the role of other counterterms in more detail.)

\vspace{4ex}

\noindent {\it 4.} Unitarity and Lorentz invariance are
consistency requirements for the continuum limit
of any lattice gauge theory. 
In the gauge-fixing approach, the restoration of Lorentz invariance
is expected to occur in a fairly standard fashion. As for
unitarity, or exact decoupling of unphysical states, this may be achieved
by imposing BRST invariance in the continuum limit.
 In fact, some of the counterterms needed to restore BRST invariance
are also needed for the restoration of Lorentz symmetry~\cite{roma}.

For this program to succeed, BRST invariance needs not
necessarily be present at finite lattice spacing. 
This observation plays a key role in the gauge-fixing approach.
In view of Neuberger's theorem~\cite{hnogo}, 
{\it not} having BRST invariance is essential for
the very existence of the lattice theory, and, hence, also for the existence of
the continuous phase transition where one can make contact with 
the target gauge-fixed continuum theory.
(In a {\it chiral} lattice gauge theory, BRST (or gauge) invariance
is  broken anyway by the fermion action. Sometimes the hope is 
expressed that this would be enough to avoid the consequences of
Neuberger's theorem. We believe that one should first formulate
gauge-fixed lattice theories without matter fields. 
If, before the introduction of matter fields, 
a gauge-fixed lattice model is ill-defined
due to exact BRST invariance, we  see little reason why the 
attempt to incorporate chiral fermions should improve the situation!)

As discussed in this paper, in the abelian case it is  appropriate
to choose a free, decoupled, lattice ghost action.
(Note that we could have chosen a ghost action for the abelian case
which is
not free on the lattice (but only in the classical continuum limit),
but there is no reason to do so, since there is no BRST invariance
on the lattice anyway.)
Now, all properties of the gauge-fixing term\seneq{Sgf} listed in
Sect.~3 generalize to the nonabelian case~\cite{mgys}. 
But in the nonabelian case we must also include a 
ghost -- gauge field  {\it interaction} term in the lattice 
action~\cite{roma,sml},
because this interaction is present in the target gauge-fixed continuum theory. 
(Note that a nonabelian ghost action {\it \`a-la} eq.~(\ref{opr}) 
will  again not have the correct classical continuum limit, and therefore
will not  be a candidate for the lattice ghost action.)

In the nonabelian  case, the measure defined 
using the Faddeev-Popov determinant
(rather than its absolute value~\cite{gribov}) is no longer positive.
Therefore, 
a possibility that one should worry about is that Neuberger's theorem still
applies in the continuum limit: approximate cancellations associated with
``smooth" continuum Gribov copies might take place, and lead to the
vanishing of the partition function in the continuum limit, even if 
such cancellations do not occur at finite lattice spacing.
Also the (related, but separate) issue of enforcing BRST invariance
nonperturbatively is highly nontrivial. These questions have to be addressed
before the gauge-fixing approach can be successfully extended to nonabelian 
theories.

\vspace{4ex}

\noindent {\em Acknowledgements}.
We thank Jan Smit for  constructive criticism. 
WB is supported by the Deutsche
Forschungsgemeinschaft under grant Wo 389/3-2, MG by
the US Department of Energy as an Outstanding Junior Investigator,
and YS by the US-Israel Binational Science
Foundation, and the Israel Academy of Science.

\vspace{2ex}


\end{document}